\documentclass[%
mathleft,%
]{an}
\usepackage{graphicx}
\usepackage{times}
\begin{document}

\Pagespan{1}{}
\Yearpublication{2011}%
\Yearsubmission{2011}%
\Month{1}%
\Volume{999}%
\Issue{92}%

\title{Fossil shell emission in dying radio loud AGNs}

\author{M. Kino\inst{1}\fnmsep\thanks{Corresponding author:
  \email{kino@kasi.re.kr}},
H. Ito\inst{2},
N. Kawakatu\inst{3},
M. Orienti\inst{4},
H. Nagai\inst{5},
K. Wajima\inst{1},
\and  R. Itoh\inst{6}
}
\titlerunning{Instructions for authors}
\authorrunning{M. Kino et al.}
\institute{
KASI,
776 Daedeokdae-ro, Yuseong-gu, 
Daejeon 305-348, Republic of Korea
\and 
Astrophysical Big Bang Laboratory, RIKEN, Saitama 351-0198, Japan
\and 
National Institute of Technology, Kure College, 
2-2-11 Agaminami, Kure, Hiroshima, 737-8506, Japan
\and
INAF - Istituto di Radioastronomia, 
via Gobetti 101, 40129 Bologna, Italy
\and
National Astronomical Observatory of Japan,
2-21-1 Osawa, Mitaka, Tokyo, 181-8588, Japan
\and
Hiroshima University, Higashi-Hiroshima, Hiroshima 739-8526, Japan}

\received{2015 June 30}
\accepted{2015}
\publonline{2015}


\abstract{
 We  investigate shell emission
associated with dying radio loud AGNs.
First,  based on our recent  work by Ito et al. (2015),
we describe the dynamical and spectral 
evolutions of shells after stopping the jet energy 
injection.
We find that the shell emission overwhelms 
that of the radio lobes soon after stopping the jet energy 
injection because fresh electrons are continuously
supplied into the shell via the forward shock 
while the radio lobes rapidly fade out
without jet energy injection.
We find that such fossil shells  
can be a new class of target sources for SKA telescope.
Next, we apply the model to the nearby radio source 3C84.
Then, we find that
the fossil shell emission in 3C84 is less luminous in radio band 
while it is bright in TeV $\gamma$ ray band and
it can be detectable by CTA.}

\maketitle

\section{Introduction}

Radio-loud active galactic nuclei (AGNs) are one of the most
powerful objects in the universe. 
According to the standard
picture of jets in AGNs,  jets are enveloped in a cocoon
consisting of shocked jet material and the cocoon is surrounded
by shocked interstellar medium region (e.g., Begelman et al. 1984). 
The shocked  interstellar medium region
(hereafter we refer to as the shell) is identical to the forward
shocked region and it is a fundamental ingredient in 
the overall AGN jet system. 
Physical properties of shells, however,
have not been well understood since they are not bright and
remain undetected (but see Croston et al. (2009) for X-ray emission from shell in Centaurus A). 
In the radio band, shells
are radio quiet (Carilli et al. 1988) and their emission is fully
overwhelmed by the radio bubbles (radio lobes).
The existence of the radio quiet shell (bow shock) is 
only identified as the discontinuity of the rotation measure
in the radio lobes of Cygnus A (Carilli et al. 1988).

In terms of observing shells, bright radio lobes prevent their detection.
Then, we come up with a question of
"what happens when jet energy injection stops?"
Motivated by this question,
we recently explore the emission from radio sources in which jet activity has ceased
at early stage of their evolution in Ito et al. (2015) (hereafter I15). 
In particular, we focus on the evolution of the relative contribution 
of the lobe and shell emission. 
In this work, we will show that
the shell will be dominant at most of the frequencies from
radio up to TeV $\gamma$-ray soon after the jet has switched off.

The layout of this paper is as follows.
In \S 2, we present the review of the model
following the formulation shown in I15.
In \S 3, we apply the model to 
a typical dying radio-loud AGN at $z=1$.
We will show that 
their fossil shell emission can be probed 
with the Square Kilometer Array (SKA).
In \S 4, we apply the model to the dying radio 
lobes observed in 3C84.
Summary and discussions are presented in \S 5.

\section{Basic Model}

Since the details of the model of pressure-driven 
expanding jet-remnant  system have been already 
well explained in Ito et al. (2011) and I15 and references therein,
here we briefly summarize
the main ingredients.

\subsection{Dynamics}

\begin{figure}
\includegraphics
[width=\linewidth]{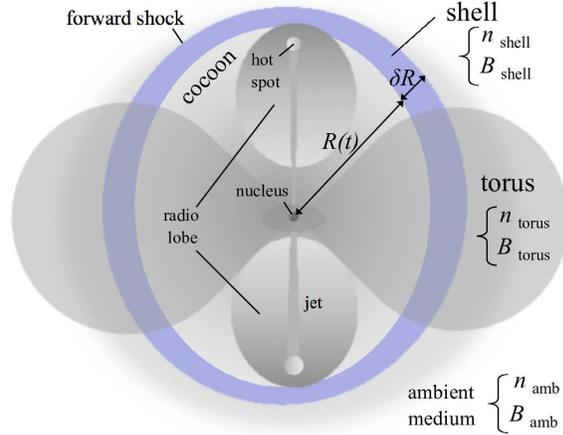}
\caption{Schematic picture of the jet and external medium interaction
in face-on view. 
The kinetic energy of the jets is dissipated via the termination shock 
at the hot spots and deposited
into the cocoon with its radius $R$ and the shell with 
its width $\delta R$. 
The cocoon is inflated by its internal energy.
The cocoon drives the forward shock propagating 
in the external medium with 
its density and magnetic field strength $n_{\rm ext}$ and $B_{\rm ext}$.
The forward shocked region is identical to the shell.}
\label{cartoon}
\end{figure}

Fig.~\ref{cartoon} presents
a general picture of a jet and external medium interaction
and we adopt the well established
expanding spherical bubble model (I15 and reference therein).
The shell width at the bubble radius ($R(t)$)
at the time $t$  is denoted by $\delta R$. 
The mass density of the
ambient matter at $R(t)$ is defined as 
$\rho(R)=\rho_{0} R$
 where $R_{0}$ is the reference radius, 
$\delta R$ satisfies the relation 
\begin{eqnarray}
\delta R= 
(\hat{\gamma}_{\rm ext}-1)[(\hat{\gamma}_{\rm ext}+1)(3-\alpha)]^{-1} R ,
\end{eqnarray}
where 
$\hat{\gamma}_{\rm ext}$ and $\alpha$ are, respectively,
the specific heat ratio of the external medium and 
the index of external ambient matter density distribution
defined as $\rho\propto R^{-\alpha}$.
We  assume that the
kinetic power of the jet ($L_{\rm j}$) is constant in time. 
The jet kinetic
energy is dissipated and deposited as the internal energy of
the cocoon and shell. 
\footnote{
The dense thermal gas surrounding
3C84 has been discovered by O'Dea et al. (1984).
However, in this work,  we focus on the diffuse ambient matter and neglect 
the dense thermal (torus) component merely for simplicity. 
The case of shocks in the dense thermal gas will be 
presented in  a forthcoming paper (Kino et al. in prep).}

In this work,
we consider two phases depending on the source age ($t$):
\begin{itemize}
\item
(i) the phase in which the jet energy injection 
into the cocoon continues
($t<t_{\rm j}$)

\item
(ii) the  phase in which the jet energy injection 
into the cocoon stops
$(t>t_{\rm j})$
\end{itemize}
where
$t_{\rm j}$
denotes the duration of the jet injection.

As for the early phase with jet energy injection
into the cocoon, 
the radius of the cocoon  is given by
\begin{eqnarray}
R(t)=C R_{0}^{\frac{\alpha}{\alpha-5}}
\left(\frac{L_{j}}{\rho_{0}}\right)^{\frac{1}{5-\alpha}}
t^{\frac{3}{5-\alpha}} \quad (t<t_{\rm j})
\end{eqnarray}
where $C$ and $R_{0}$ are 
the numerical coefficient and 
the reference size of the cocoon,
respectively.
Note that 
$L_{\rm j}/\rho_{0}$ is the key quantity
which controls the dynamical
evolution of AGN jet system
(e.g., Kawakatu et al. 2009).

After the energy injection
from the jet ceases, the cocoon will rapidly lose its energy due to adiabatic expansion and
most of its energy is transferred  into the shell within a dynamical timescale.
Hence, after the transition time, the cocoon pressure becomes dynamically unimportant,
and the energy of the shell becomes dominant. Therefore, in the late phase, expansion of the
bow shock asymptotically would follow the Sedov-Taylor expansion. 
Therefore we set 
\begin{eqnarray}
\dot{R}(t) \propto t^{-(3-\alpha)/(5-\alpha)}
 \quad (t>t_{\rm j}) .
\end{eqnarray}
In this phase, 
the adiabatic relation $P_{c}V_{c}^{\hat{\gamma}}={\rm const.}$
should hold.
Then, we can approximately describe  the late phase as follows:
\begin{eqnarray}
R(t)=
R(t_{\rm j})
\left[\frac{P(t_{\rm j})}{P(t)}\right]^{1/3\hat{\gamma}}
 \quad (t>t_{\rm j})  .
\end{eqnarray}
We set 
$\hat{\gamma}_{\rm ext}=5/3$ and
$\hat{\gamma}=4/3$
for the shock jump condition between
the shell and external medium (I15).

\subsection{Non-thermal emission}

Since the details of the treatment of non-thermal emissions
have been already explained in Kino et al. (2013) (hereafter K13)
and I15,
we briefly show the basic treatment of photon and electron
distributions in shells and lobes. 
We solve the set of kinetic
equations describing the electron and photon 
energy distributions.
First, 
as for the external photon field against inverse Compton (IC) process, 
(1) UV photons from a standard accretion disk, 
(2) IR photons from a dust torus, 
(3) synchrotron photons from the radio lobes, and 
(4) synchrotron photons from the shell
are included.
Second,
we include the effect of absorption via $\gamma \gamma$  interaction.
VHE photons suffer from  absorption via interaction
with various soft photons. 
Here, we include the $\gamma \gamma$  absorption 
due to both source-intrinsic and EBL (Extragalactic Background Light)
photon fields. The absorption opacity with respect
to the intrinsic photons can be calculated by summing
up all of the photons from (1) the shell, (2) the radio lobes,
(3) the dusty torus, and (4) the accretion disk and we multiply
the  $\gamma \gamma$ absorption factor of $\exp(-\tau_{\gamma \gamma})$ 
with the unabsorbed flux where $\tau_{\gamma \gamma}$ is the 
optical depth for the  $\gamma \gamma$ absorption.
For simplicity, we deal with the absorption effect at the first
order and we neglect cascading effect. With regard to the opacity
for  $\gamma \gamma$ interaction between 
EBL
and TeV $\gamma$ photons, we adopt
the standard model of Franceschini et al. (2008).

\section{Application to high-z dying radio sources}

\subsection{Setting}

Here we demonstrate the time evolutions of the energy 
distribution of non-thermal electrons and the resulting emission. 
As a fiducial case we focus on sources
with jet power of $L_{\rm j} = 1\times 10^{45}~{\rm erg~s^{-1}}$
at $z=1$. 
We set the duration of energy
injection as $t_{\rm j} = 10^{5}$~yr. The chosen value of the age is
around the upper end of the estimated age of the compact radio source 
with an overall linear size of 5 kpc 
(e.g.,  Murgia et al. 1999).

Regarding the IC scattering, full Klein-Nishina (KN)
cross section is taken into account. 
As a source of seed photons, 
we consider UV emission from the accretion disc, IR emission
from the dusty torus, stellar emission in NIR from the
host galaxy, synchrotron emission from the radio lobe
and cosmic microwave background (CMB). We constructed the model of
the spectra of the photons from the disc, torus and host
galaxy with a black-body spectra.
As for the luminosities of the emissions, we adopt 
$L_{\rm UV}=L_{\rm IR}=1\times 10^{45}~{\rm erg~s^{-1}}$
for the disc and torus emission (e.g., Elvis et al. 1994;
Jiang et al. 2006), and 
$L_{\rm IR}=1\times 10^{44}~{\rm erg~s^{-1}}$
for the host galaxy emission (e.g., de Ruiter et al. 2005).

\subsection{Results: Long-lived shell emission 
in high-z radio-loud sources as a new target for SKA}

\begin{figure}
\includegraphics[width=\linewidth]{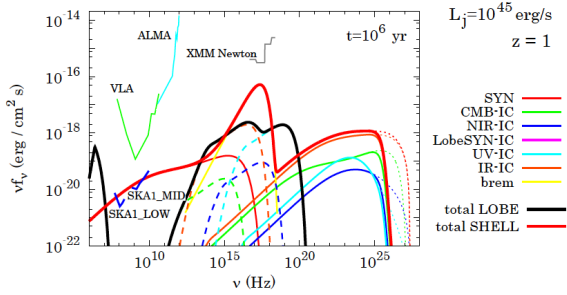}
\caption{
Broadband spectrum of a dying radio source 
with a jet power of
$ L_{\rm j} = 1\times 10^{45}~{\rm erg~s^{-1}}$
located at z = 1.
The figure is adopted from I15.
The source age in this case is $10^{6}$ yr 
and the jet injection has been stopped since $t_{\rm j}=10^{5}$ yr. 
The thick lines show the total 
flux from the shell and lobe. 
The various thin dashed and thin solid 
lines show the contributions to the total emission produced in the lobe
and shell, respectively.
(A color version of this figure is available in the online journal.)
The shell emission can be detected by SKA.
The SKA sensitivity curves correspond to 3 sigma detection 
limit with 10 hours of integration time 
(Prandoni \& Seymour 2015).}
\label{ska}
\end{figure}

In Fig.~\ref{ska}, we show the resultant spectra from 
the radio lobe and shell after the jet injection was stopped.
We examine the typical case with the conservative
jet power
$ L_{\rm j} = 1\times 10^{45}~{\rm erg~s^{-1}}$ at $z=1$,
and the electron acceleration efficiency 
in the shell $\epsilon_{e,\rm shell}=0.01$.
We found that the shell emission overwhelms 
that of the radio lobes soon after stopping the jet energy 
injection because fresh electrons are continuously
supplied into the shell via the forward shock 
while the radio lobes rapidly fade out.

Regarding the detectability of the shell emission, 
SKA telescope is capable of 
detecting the emission. The detection is
marginal for SKA phase 1.
Since these values inevitably have dispersion from source
to source, we expect a certain fraction of sources to
be well above the detection limit. Moreover, in the phase
2 (SKA2), the sensitivity is expected to improve by an order 
of magnitude. Hence, SKA will be a powerful tool
to reveal the population of dead radio sources which are
dominated by the shell emission.

\section{Application to 3C84}

The compact radio source 3C84,
also known as the Seyfert galaxy NGC1275
(at the redshift $z=0.0176$)
is one of the ideal target for this study.

\begin{figure}
\includegraphics[width=\linewidth]{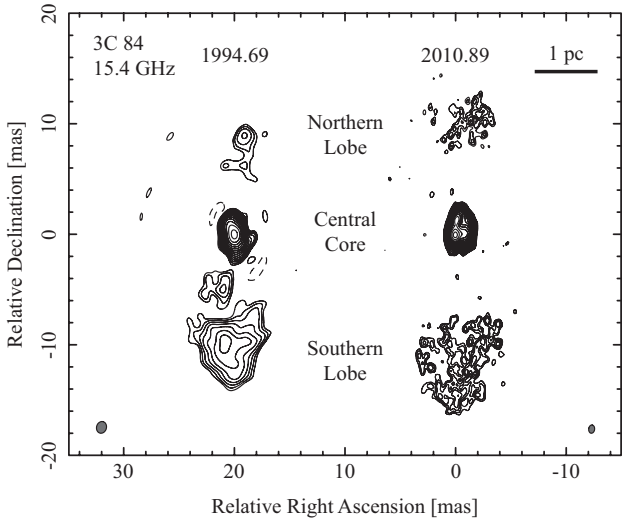}
\caption{
VLBA images of 3C 84 obtained at 15.4 GHz. 
Images refer to observations performed 
on 1994 September 10 (left) and 2010 November 20 (right). 
The lowest contour
is three times the off-source rms noise (one $\sigma$), and the contour levels
are $-3\sigma$, $3\sigma \times (\sqrt{2})^n$ ($n$ = 0, 1, 2, ...). The peak
intensity is 5.86 Jy~beam$^{-1}$ (left) and 5.54 Jy~beam$^{-1}$ (right). 
The restoring beam is indicated
in the bottom left/right corner of the left/right images.
The total flux density of northern  and southern lobes are 0.8~Jy and 6.7~Jy in 1994,
while they decrease to 0.5~Jy and 1.4~Jy in 2010, respectively.}
\label{15G}
\end{figure}

In Fig.~\ref{15G}, we show the two epoch comparison of 
VLBA images  obtained  at 15~GHz
obtained in 1994  and 2010. 
From this, we can clearly see that the pair of outer radio lobes
on $\sim10$~mas scale is fading out due to the lack of
jet energy injection.
Hereafter, we examine the shell emission 
associated with this fading radio lobes.

\subsection{Setting}

Number densities of surrounding medium are 
important quantities but they have some degree of  uncertainties.
Taylor et al. (2006) estimated the number density
by using deep Chandra observation (Fabian et al. 2006).
Within the central 0.8~kpc,
the density profile is severely affected by
the nucleus. So, they estimated an average
central density over the inner 2~kpc to be
 $n_{\rm amb}\sim 0.3~{\rm cm^{-3}}$.
Regarding the upper limit on the number density,
recently Fujita et al. (2014) obtained
$n_{\rm amb}\sim 10~{\rm cm^{-3}}$
for the inner part of Perseus cluster
by  taking the following two  assumptions:
(1) hot gas outside the Bondi radius is in nearly a hydrostatic
equilibrium in a gravitational potential, and 
(2) the gas temperature near the galaxy
centre is close to the virial temperature of the galaxy.
In this work, we adopt 
$n_{\rm amb}=1~{\rm cm^{-3}}$
which is in between these values.

Energy densities of surrounding photon fields are also 
important quantities. They are seed photons for IC scattering.
As for UV from the accretion disc,
we set
\begin{eqnarray}
　 L_{UV} = 5\times 10^{42}~ {\rm erg~s^{-1}} ,
\end{eqnarray}
based on the observation of 
Kanata telescope (see Appendix)
and we conservatively include the safety factor $1/2$ to
mimic a possible contamination of extended sources.
The accretion flow can be safely regarded as a point source.
Regarding IR dust, we set
$L_{\rm IR}=L_{\rm UV}/2$
based on Calderone et al. (2012).
Note that IR torus is supposed to be extended up to 10~pc scale
in order to produce the free-free absorption observed at  
the northern radio lobe (Walker et al. 2000).
Fluxes of synchrotron emission from radio lobes are 
adopted from the actual observational data.
As for the mini radio lobes at $\sim3$~mas scale
(Nagai et al. 2010; Nagai et al. 2012; Suzuki et al. 2012), we set
\begin{eqnarray}
　L_{\rm mini-lobe} \approx 1\times 10^{43}~{\rm erg~s^{-1}}
\end{eqnarray}
at $\sim 200$~GHz based on the recent flux level of 3C84 
in SMA calibrator list.

Let us discuss the total jet power  in 3C84.
The observed luminosity at each energy band is 
of order of $\sim 10^{43}~{\rm erg~s^{-1}}$
from radio to GeV $\gamma$-ray band (Abdo et al. 2009).
Therefore, the bolometric luminosity is estimated 
as close to $\sim 10^{44}~{\rm erg~s^{-1}}$.
Hence, for example, a typical  $10~\%$
radiative efficiency of non thermal electrons
results in the electron kinetic power 
of order of close to $\sim 10^{45}~{\rm erg~s^{-1}}$.
Proton component is also supposed to 
contribute to the total jet power.
Based on this order estimation,
we set  $L_{\rm j}=3\times 10^{45}~{\rm erg~s^{-1}}$
 in the present work.

\subsection{Results: Long-lived shell emission in 3C84 as a new target of CTA}

\begin{figure}
\includegraphics
[width=\linewidth]{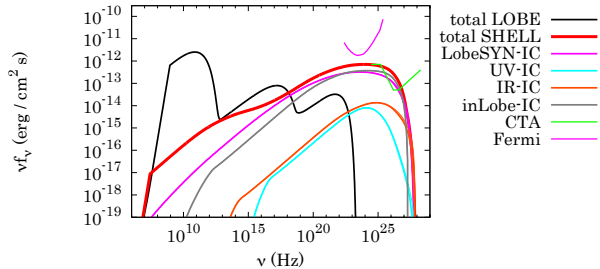}
\caption{
Fossil shell spectrum
10 years after the jet cessation in 3C84 
(the thick curve peaked at gamma-ray energy band). 
The shell spectrum is IC dominated.
There are four components of the seed photons 
for IC scattering;
i.e., 
synchrotron photons from outer and inner radio lobes, 
UV from the accretion disk, and
IR from the torus.
Among them, synchrotron photons 
from outer and inner radio lobes are dominant
(A color version of this figure is available in the online journal.)}
\label{low_n}
\end{figure}

In Fig. \ref{low_n}, we show a prediction of
the  broadband spectrum from the fossil shell in 3C84
10 years after the jet cessation.
The significant difference between this case and 
the above section is  the distance of the sources from the Earth. 
Since the redshift of 3C84 is sufficiently low, 
$\gamma \gamma$ absorption with the EBL is less effective.
Therefore, it can be a  new TeV gamma-ray emitter candidate.
We find that the IC emission can be detectable by 
the Cherenkov Telescope Array (CTA) in this case.
We also find that the synchrotron photons from the
radio lobes are the dominant seed photons for the IC
scattering in 3C84.
The detailed comparison of the contribution 
from mini-radio lobes will be presented in 
a forthcoming paper (Kino et al. in preparation).


\begin{table}
\caption{Parameters of radio lobes and shell in 3C84}
\label{tlab}
\begin{tabular}{ccccc}\hline
$L_{\rm j}$ & $B_{\rm shell}$ & $n_{\rm amb}$ &  $\epsilon_{e, \rm shell}$ & $\epsilon_{e, \rm lobe}$\\ 
(erg~s$^{-1}$) & (mG) & (${\rm cm^{-3}}$) &-- & -- \\
\hline
$3\times 10^{45}$  & 0.2 & 1 & 0.1 & 0.1 \\
\hline
\end{tabular}
\end{table}

\section{Summary}

In the present work
we studied a possible fossil shell emissions
associated with dying radio loud AGNs.
Below, we summarize the main results.

\begin{itemize}

\item[--]

We reviewed our recent work  presented in I15.
We have examined 
the dynamical and spectral 
evolution of fossil shells 
which are identical to the forward shocks
propagating in the external medium.
We find that the shell emission overwhelms 
that of the radio lobes soon after stopping the jet energy 
injection because fresh electrons are continuously
supplied into the shell via the forward shock 
while the radio lobes rapidly fade out.

\item[--]

Fossil shell emission in 
high-z dying radio sources is presented.
We examine the typical case with the conservative
jet power $L_{\rm j}=1\times 10^{45}~{\rm erg~s^{-1}}$ at $z=1$,
and electron acceleration efficiency
in the shell $\epsilon_{e,\rm shell}=0.01$.
Regarding  future facilities, SKA telescope is capable of detecting the emission
in the radio band. The detection is
marginal for SKA phase 1.

\item[--]

Fossil shell emission in 3C84  
is investigated by applying I15 model.
Since 3C84 is a low redshift source, 
$\gamma \gamma$ absorption is less effective.
We examine 
the  broadband spectrum from the fossil shell in 3C84
10 years after the jet cessation
with the jet power $L_{\rm j}=3\times 10^{45}~{\rm erg~s^{-1}}$,
and relatively large electron acceleration efficiency
in the shell $\epsilon_{e,\rm shell}=0.1$.
We find that the IC emission can be detectable by 
CTA.

\end{itemize}


\acknowledgements
  
 The research leading to these results has received funding from
the  European Commission Seventh Framework Programme (FP/2007-2013)
under grant agreement No 283393 (RadioNet3).
Part of this work was done with the contribution of the Italian Ministry
of Foreign Affairs and University and Research for the
collaboration project between Italy and Japan.
This research has made use of data from the MOJAVE database that is 
maintained by the MOJAVE team (Lister et al., 2009) .

%

\begin{thebibliography}{}




\bibitem[Abdo et al.(2009)]{2009ApJ...699...31A} Abdo, A.~A., Ackermann, 
M., Ajello, M., et al.\ 2009, \apj, 699, 31 

\bibitem[Begelman et al.(1984)]{1984RvMP...56..255B} Begelman, M.~C., 
Blandford, R.~D., \& Rees, M.~J.\ 1984, Reviews of Modern Physics, 56, 255 


\bibitem[Calderone et al.(2012)]{2012MNRAS.425L..41C} Calderone, G., 
Sbarrato, T., \& Ghisellini, G.\ 2012, \mnras, 425, L41 


\bibitem[Carilli et al.(1988)]{1988ApJ...334L..73C} Carilli, C.~L., Perley, 
R.~A., \& Dreher, J.~H.\ 1988, \apjl, 334, L73 

\bibitem[Croston et al.(2009)]{2009MNRAS.395.1999C} Croston, J.~H., Kraft, 
R.~P., Hardcastle, M.~J., et al.\ 2009, \mnras, 395, 1999 


\bibitem[de Ruiter et 
al.(2005)]{2005A&A...439..487D} de Ruiter, H.~R., Parma, P., Capetti, A., et al.\ 2005, A\&A, 439, 487 

\bibitem[Elvis et al.(1994)]{1994ApJS...95....1E} Elvis, M., Wilkes, B.~J., 
McDowell, J.~C., et al.\ 1994, \apjs, 95, 1 


\bibitem[Fabian et al.(2006)]{2006MNRAS.366..417F} Fabian, A.~C., Sanders, 
J.~S., Taylor, G.~B., et al.\ 2006, \mnras, 366, 417 


\bibitem[Franceschini et 
al.(2008)]{2008A&A...487..837F} Franceschini, A., Rodighiero, G., \& Vaccari, M.\ 2008, A\&A, 487, 837 


\bibitem[Fujita et al.(2014)]{2014arXiv1406.6366F} Fujita, Y., Kawakatu, 
N., \& Shlosman, I.\ 2014, arXiv:1406.6366 


\bibitem[Greene 
\& Ho(2005)]{2005ApJ...630..122G} Greene, J.~E., \& Ho, L.~C.\ 2005, \apj, 630, 122 


\bibitem[Ho et al.(1997)]{1997ApJS..112..391H} Ho, L.~C., Filippenko, 
A.~V., Sargent, W.~L.~W., \& Peng, C.~Y.\ 1997, \apjs, 112, 391 

\bibitem[Ito et al.(2011)]{2011ApJ...730..120I} Ito, H., Kino, M., 
Kawakatu, N., \& Yamada, S.\ 2011, \apj, 730, 120 

\bibitem[Ito et al.(2015)]{} Ito, H., Kino, M., 
Kawakatu, N., \& Orienti, M.\ 2015, \apj, in press (I15)


\bibitem[Jiang et al.(2006)]{2006AJ....132.2127J} Jiang, L., Fan, X., 
Hines, D.~C., et al.\ 2006, \aj, 132, 2127 


\bibitem[Kawabata et al.(2008)]{2008SPIE.7014E..4LK} Kawabata, K.~S., 
Nagae, O., Chiyonobu, S., et al.\ 2008, SPIE, 7014, 70144L 

\bibitem[Kawakatu et al.(2009)]{2009ApJ...697L.173K} Kawakatu, N., Kino, 
M., \& Nagai, H.\ 2009, \apjl, 697, L173 


\bibitem[Kino et al.(2013)]{2013ApJ...764..134K} Kino, M., Ito, H., 
Kawakatu, N., \& Orienti, M.\ 2013, \apj, 764, 134 


 \bibitem[Lister et al.(2009)]{2009AJ....137.3718L} Lister, M.~L., Aller, 
H.~D., Aller, M.~F., et al.\ 2009, \aj, 137, 3718 

 \bibitem[Murgia et 
al.(1999)]{1999A&A...345..769M} Murgia, M., Fanti, C., Fanti, R., et al.\ 1999, A\&A, 345, 769 

 
 \bibitem[Nagai et al.(2012)]{2012MNRAS.423L.122N} Nagai, H., Orienti, M., 
Kino, M., et al.\ 2012, \mnras, 423, L122 

\bibitem[Nagai et al.(2010)]{2010PASJ...62L..11N} Nagai, H., Suzuki, K., 
Asada, K., et al.\ 2010, \pasj, 62, L11 



\bibitem[O'Dea et al.(1984)]{1984ApJ...278...89O} O'Dea, C.~P., Dent, 
W.~A., \& Balonek, T.~J.\ 1984, \apj, 278, 89 



\bibitem[Prandoni 
\& Seymour(2015)]{2015aska.confE..67P} Prandoni, I., \& Seymour, N.\ 2015, 
Advancing Astrophysics with the Square Kilometre Array (AASKA14), 67 



\bibitem[Suzuki et al.(2012)]{2012ApJ...746..140S} Suzuki, K., Nagai, H., 
Kino, M., et al.\ 2012, \apj, 746, 140 

\bibitem[Taylor et al.(2006)]{2006MNRAS.368.1500T} Taylor, G.~B., 
Gugliucci, N.~E., Fabian, A.~C., et al.\ 2006, \mnras, 368, 1500 



\bibitem[Walker et al.(2000)]{2000ApJ...530..233W} Walker, R.~C., Dhawan, 
V., Romney, J.~D., Kellermann, K.~I., 
\& Vermeulen, R.~C.\ 2000, \apj, 530, 233 


\bibitem[Yamazaki et al.(2013)]{2013PASJ...65...30Y} Yamazaki, S., 
Fukazawa, Y., Sasada, M., et al.\ 2013, \pasj, 65, 30 



 
\end{thebibliography}
%

\newpage

\appendix

\section{The H$\alpha$ line luminosity in NGC~1275}

\begin{figure}
\includegraphics
[width=\linewidth]
{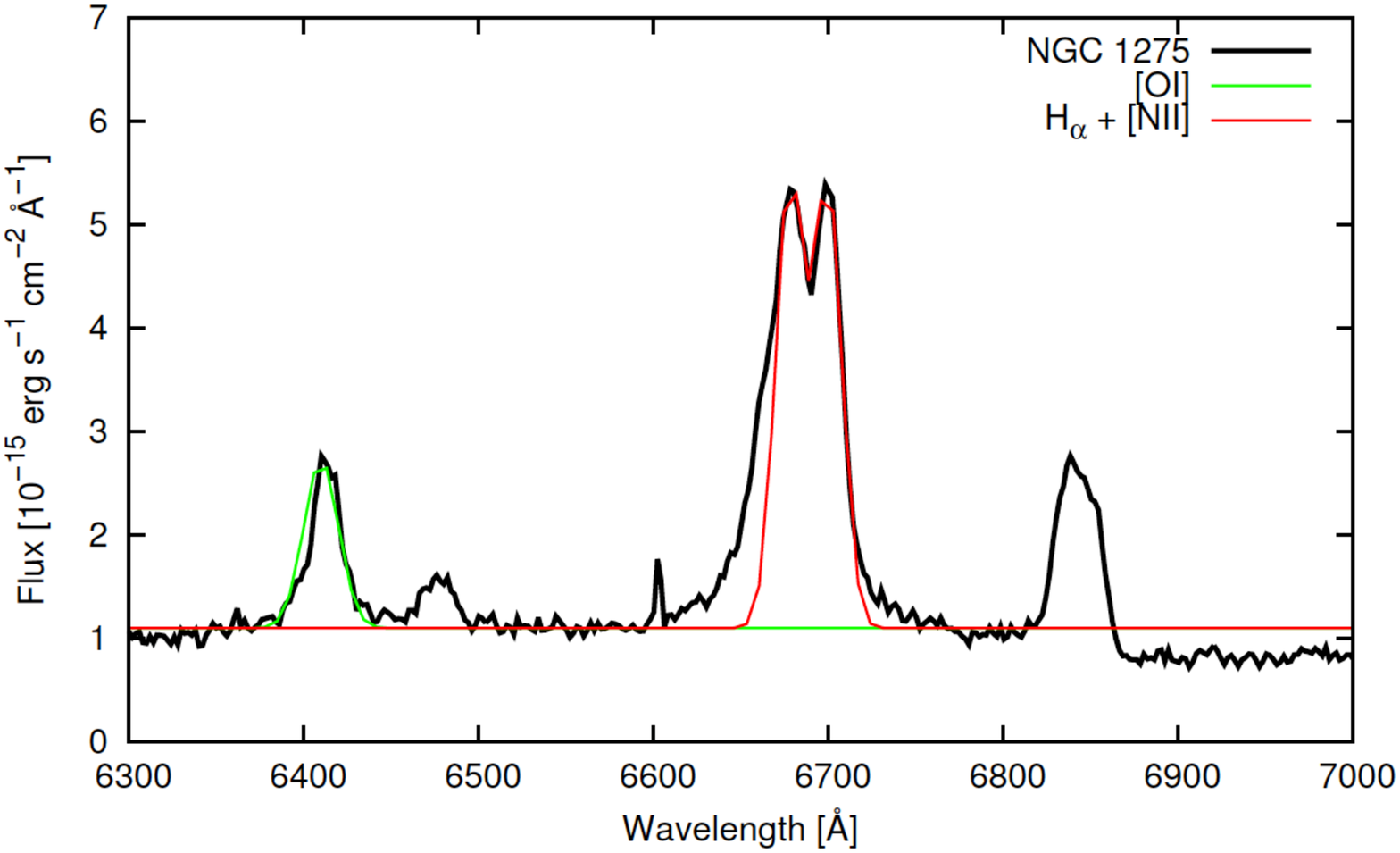}
\caption{Overall line spectra (the thick line)
for the nucleus of NGC~1275 measured by HOWPol on the
Kanata telescope.
The [OI] flux (the line at $\sim$ 6400~\AA)
is scaled to the value reported in Ho et al. 1997.
The sum of H$\alpha$ (both broad and narrow) 
and [NII] lines are shown by the red line.
We note that there is an ambiguity due to the unclear
Hubble constant value adopted in Ho et al. 1997. }
\label{HOWPol}
\end{figure}

Here we evaluate the 
 H$\alpha$ luminosity ($L_{\rm H\alpha}$) of NGC~1275 using 
 the observation data obtained by Kanata telescope's
 HOWPol (Hiroshima One-shot Wide-field Polarimeter)
(Kawabata et al. 2008).
We revisit the  data used in Yamazaki et al. (2013)
 since 
(1) the flux  was integrated over an aperture of $4''.7$ width, 
and
(2) [NII] and narrow H$\alpha$  lines were not taken into account,
which may lead to an overestimate of $L_{\rm H\alpha}$.
In Fig.~\ref{HOWPol} we show the optical spectrum at the nucleus of NGC~1275.
The flux values of [OI] and narrow line of H$\alpha$ 
are normalized to the literature values shown in Ho et al. (1997).
The broad H$\alpha$ 
and [NII] lines are adjusted to fit the observed data
by eye inspection.
 Then we obtain that
 $L_{\rm H\alpha, broad}\sim 3.2\times 10^{41}~{\rm erg~s^{-1}}$.
 Then we can derive the continuum luminosity 
 \begin{eqnarray}
 L_{5100}\sim 1\times 10^{43}~{\rm erg~s^{-1}},
 \end{eqnarray}
by using the well known empirical relation 
Greene and Ho (2005).
Although there may include some contaminations from
extended components (e.g., nebula, filaments etc), 
we use this value as a best estimate for $L_{\rm H\alpha}$.
The obtained $L_{\rm H\alpha}$ is about three times
smaller than that in Yamazaki et al. (2013).

\end{document}